\documentclass[conference]{IEEEtran}
\IEEEoverridecommandlockouts
\usepackage{cite}
\usepackage{amsmath,amssymb,amsfonts}
\usepackage{graphicx}
\usepackage{textcomp}
\usepackage{xcolor}

\usepackage{algpseudocode}
\usepackage{algorithm}
\def\BibTeX{{\rm B\kern-.05em{\sc i\kern-.025em b}\kern-.08em
    T\kern-.1667em\lower.7ex\hbox{E}\kern-.125emX}}
\begin{document}

\title{Towards AI-enabled Control for Enhancing Quantum Transduction}

\author{\IEEEauthorblockN{Mekena Metcalf}
\IEEEauthorblockA{\textit{Computational Research Division} \\
\textit{Lawrence Berkeley National Laboratory}\\
Berkeley, CA \\
mmetcalf@lbl.gov}
\and
\IEEEauthorblockN{Anastasiia Butko}
\IEEEauthorblockA{\textit{Computational Research Division} \\
\textit{Lawrence Berkeley National Laboratory}\\
Berkeley, CA \\
abutko@lbl.gov}
\and
\IEEEauthorblockN{Mariam Kiran}
\IEEEauthorblockA{\textit{Scientific Networking Division} \\
\textit{Lawrence Berkeley National Laboratory}\\
Berkeley, CA\\
mkiran@lbl.gov}
}

\maketitle

\begin{abstract}
With advent of quantum internet, it becomes crucial to find novel ways to connect distributed quantum testbeds and develop novel technologies and research that extend innovations in managing the qubit performance. Numerous emerging technologies are focused on quantum repeators and specialized hardware to extend the quantum distance over special-purpose channels. However, there is little work that utilizes current network technology, invested in optic technologies, to merge with quantum technologies. In this paper we argue for an AI-enabled control that allows optimized and efficient conversion between qubit and photon energies, to enable optic and quantum devices to work together. Our approach integrates AI techniques, such as deep reinforcement learning algorithms, with physical quantum transducer to inform real-time conversion between the two wavelengths. Learning from simulated environment, the trained AI-enabled transducer will lead to optimal quantum transduction to maximize the qubit lifetime. 
\end{abstract}


\section{Introduction}
As qubit technology advances, Noisy Intermediate Scale Quantum (NISQ) devices need to be scaled to meet growing computation demands. This is mainly caused by the physical limitations that suppress the number of qubits that can fit on a single chip. These limitations can be addressed by entangling multiple quantum chips with high fidelity, particularly using optical fibers. A successful technology capable of doing this would enable multi-nodal quantum systems and quantum networks connecting largely separated quantum facilities similar to the \textit{quantum internet} vision \cite{osti_1638794}.  

Due to the nature of quantum computing, quantum information cannot be copied between devices (\textit{no-cloning theorem} \cite{1970Park}), but it can be converted into different forms and moved into a different location. However, connecting quantum devices via an optical network requires an ability to convert a qubit on a quantum device into a photonic or \textit{flying qubit}. This process of the qubit conversion is called \textit{quantum transduction} and is usually done by a physical converter (\textit{transducer}). There are several transduction approaches that differ in their technologies, conversion bandwidth and efficiency, e.g. optomechanics, electro-optics, Rydberg atoms ensemble and more \cite{QTRev2020}. Among others, optical-to-microwave transduction has the potential to achieve best-to-date conversion efficiency. 

The conversion between microwave (qubit) and optical (network) frequencies is an extremely challenging task. It is severely affected by disparate energy scales, noise, decoherence, cavity cooperativity and bandwidth characteristics, which can all drop the conversion efficiency below the level of usefulness. 
Among existing implementations, \textit{there is still no experimental demonstration of practical quantum transduction with conversion efficiency at the quantum threshold - the bare minimum for a converter to be useful}.  

A dominant source of low conversion efficiency for quantum transduction stems from thermal noise in the mechanical converter. 
Without a method to mitigate thermal noise from the converter, most experiments have conversion efficiencies below $10\%$~\cite{QTRev2020}. A recent experiment done by a team at National Institute of Standards and Technology (NIST)~\cite{NIST_qt} applied a feed-forward protocol to an optomechanical transducer to reduce added noise. Authors demonstrated conversion efficiency within 10\% of the quantum threshold. However, efficiency of their experiment was determined by fitting experimental data to theoretical solutions and the assumptions therein. 
Theoretical techniques to model {\it open-quantum systems} assume that the system and environment are weakly coupled, dynamics are Markovian, and information is lost in an irreversible process. While in experiment, the dynamics may also be non-Markovian which results in dephasing errors, however, the information remains recoverable from the state-space of the system~\cite{QNL_BackAction}.

To overcome these challenges, we argue for a radically novel approach to solve the \emph{quantum transduction problem using Artificial Intelligence (AI) by performing data-driven learning for optimal conversion efficiency between optic and microwave frequencies}. We argue that the AI can be trained to learn effects of noise on the wave-function and recover information spread lost from dephasing. This information recovery can yield in accurate characterization of device performance and increase the transduction efficiency producing radically impactful results for the challenge. 

To enable the data-driven AI learning, we propose to use a model-free stochastic deep reinforcement learning (DRL) approach, that allows an agent to learn optimal transducer behavior by interacting with the quantum environment. To train the AI agent, we develop a \textit{transducer model} that can accurately simulate the device behaviour over time, and with advanced noise models applied, to optimize the optomechanical conversion efficiency. These techniques have been demonstrated to work in theory, but no practical implementation exists \cite{Google_RLQC}. The DRL solutions can be validated against the theoretically optimum solution used to parameterize the device in the NIST experiment.

The rest of the paper has been organized as follows. Section \ref{sec:motiv} describes the motivation and the recent studies in the area of quantum transduction. Section \ref{sec:method} describes our methodology towards AI-enabled control for enhancing quantum transduction. Section \ref{sec:sim} gives a brief overview of the simulation and training environment. In Section \ref{sec:drl}, we describe our DRL algorithm and adaptive learning approach. Finally, Section \ref{sec:conc} concludes our work discussing the potential impact of it on the field.  

\section{Motivation and Background}
\label{sec:motiv}
Complex dynamics of quantum systems are controlled using electromagnetic field temporal and spectral Degrees Of Freedom (DOF) \cite{Brief2010}. Control objectives are framed as cost-functionals that optimize the evolution control, state control or observable control using the electromagnetic field DOFs. Optimizing the control trajectories in real quantum experiments is challenging since the state space grows exponentially and the cost function may not accurately represent genuine system dynamics. Data driven machine learning techniques offer a viable approach to optimize quantum system dynamics in real experiments.

\begin{figure*}[t]
	\includegraphics[width =\textwidth]{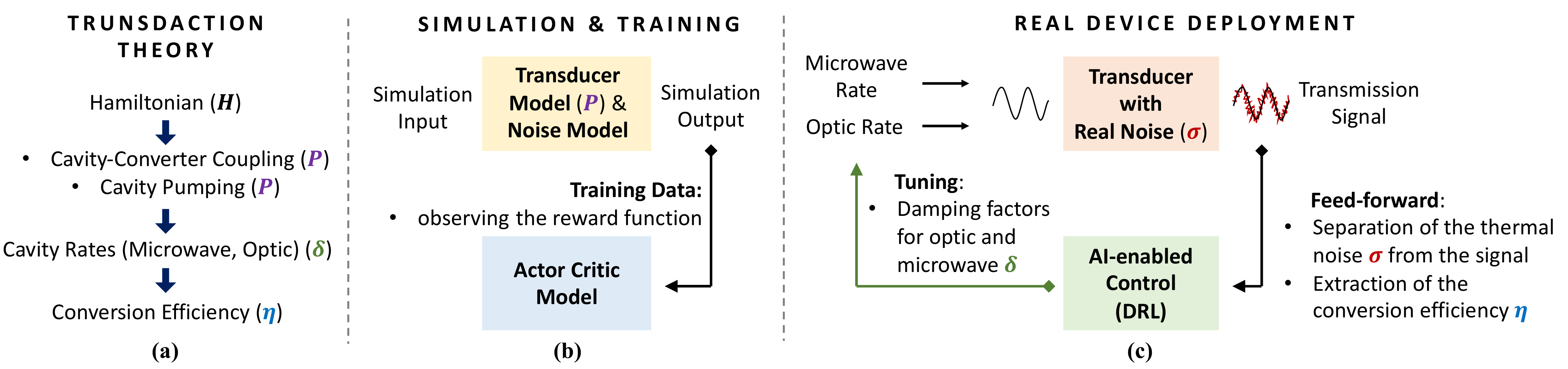}
	\caption{Overview of optomechanical quantum transduction and connection proposed DRL methodology.}
	\label{fig:methodology}
\end{figure*}

Deep Reinforcement Learning (DRL) techniques have proven to be valuable in evaluating a high-dimensional optimization problems such as game playing, robotic control and managing data centers \cite{mnih}. It builds upon classical models, replacing the learning with a neural network to approximate policy and value functions. Here, the function approximates the environment state space with actions and rewards. Particularly, when the state space is too large to store, this approach has proved feasible in learning approximate conditions. The algorithm formulates an agent that is in a partially observable environment, learning from interactions the best decisions. The agent observes environment snapshots, and chooses an action, receiving a reward value for future actions in the current state. The agent receives feedback on its action in form of rewards, until terminal state is reached. The objective is to maximize the cumulative reward over all actions in the time agent is active \cite{Sutton:2018:RLI:3312046}. 

Similar to learning optimal strategies in playing games, a DRL technique can be trained against noise to evaluate stochastic quantum processes and learn optimum actions. A recent experiment done by Google implemented a time-dependent DRL trained in a noisy environment~\cite{Google_RLQC}. When applied to optimal quantum control, it  minimizes leakage errors reducing gate error by two orders of magnitude. This experiment showed recovery of quantum information with an optimized control solution generated using time-dependent DRL techniques that can outperform the theoretically optimal solution.

In optical-to-microwave transduction, microwave and optical cavities interact by a mechanical vibrational mode to convert photonic states between the disparate frequency regimes quantified in the rotating frame by the Hamiltonian

\begin{multline}
\label{Eq:Ham}
    H = \omega_m \hat{n}_i^m +\sum_l [\omega_{c,l} \hat{n}_{l,c} + \gamma_l (\hat{b}_m^\dagger + \hat{b}_m)\hat{n}_{c,l}\\ 
     + i\epsilon_j(\hat{a}_j^\dagger e^{-i\omega_{d,j}t}-\hat{a}_j e^{i\omega_{d,j}t}) ].
\end{multline}

The first term in the Hamiltonian is the energy of the mechanical mode, the second term represents the energy of the cavity, the third term describes the interaction between the mechanical mode and the cavity with interaction strength $\gamma$, the final term represents optical pumping of the cavities with an electromagnetic field. Our initial insight is applying quantum optimal control to tailor the pumping pulses to optimize cavity efficiency. Generally, quantum optimal control is used to tailor single and two qubit quantum interactions, but we investigate the extent to which we can employ quantum optimal control methodologies to quantum network capabilities.

\section{Towards an AI Controlled End-to-End Transduction}
\label{sec:method}

Our methodology for AI-enabled control for enhancing quantum transduction is shown in Figure~\ref{fig:methodology}. It includes three major phases: (a) transduction theory, (b) AI agent training through simulation, and (c) real device deployment.

\textbf{Transduction Theory:} Seminal theoretical evaluations of the optomechanical transduction scheme made approximations to the Hamiltonian in Equation \ref{Eq:Ham}, and when both cavity detunings are equal to the mechanical mode the expression is the appropriate beam-splitter form needed for quantum transduction.
\begin{equation}
    H^\prime = \sum_l G_l (\hat{a}_l - \sqrt{n_l})\hat{b}^\dagger + (\hat{a}^\dagger_l - \sqrt{n_l})\hat{b}
\end{equation}
with $n_l$ being the mean number of photons induced by the pumps and $G_l = \gamma_l\sqrt{n_l}$ is the cavity rate \cite{TianTheory, Painter2011}. These cavity rates are realized in the NIST experiment using tunable optical pumps. We consider the cavity-converter interaction parameter $\gamma$ and the cavity pump temporal and spectral DOFs as parameters in the cost function since they control the coupling between the cavities and mechanical converter. Assuming weak coupling to the environment and making the Markov approximation, the conversion efficiency $\eta$ can be used to characterize an optomechanical transducer \cite{Lecocq2016}.
\begin{equation}
    \eta = \frac{4 C_1 C_2}{1+C_1 + C_2}
\end{equation}
\begin{equation}
    C_l = \frac{4 G_l^2}{\delta_l \delta_m}
\end{equation}
The NIST team fit experimental data to this metric. We, however, consider a different approach by designing a reward function based on the scattering matrix that can go beyond the weak coupling regime to better bridge theory and experimental data using AI.


\textbf{Simulation \& Training:} We employ a policy gradient learning approach such as the \textit{Actor-Critic Model} to allow two neural networks to learn best policy for tuning and action-value to learn optimal tuning parameters. We explore the actor-critic model, the `Critic' estimates the value function, e.g. from the action-value (the Q value), and the `Actor' updates the policy distribution in the direction suggested by the Critic (such as with policy gradients). This model enables the AI agent to learn optimal actions in a continuous action space, suitable for tuning parameters.

Training data is essential to train the agent with detailed knowledge of the device state space and optimal actions mapped. As this is a nascent field, transducer devices are not available to optimally train our agent. To address this challenge, we develop a parameterizable device simulation model, based on theory, for the transducer behaviour, incorporating noise model by emulating dynamics of non-classical photons coupled to the environment~\cite{Clerk2010}. A cost function is constructed from Hamiltonian theory by tuning the cavity rates, keeping noise parameters fixed. The DRL agent is programmed to receive a reward when conversion efficiency increases from past experiences.

Learning in the simulation will be in discrete time steps with the control loop making immediate decisions. However, our goal is to demonstrate the trained DRL to work with real world devices and we will demonstrate this by coupling with a real device and allowing the trained DRL to perform online learning to reach optimal conversion efficiency quickly.

\textbf{Real Device Deployment:} Prior research in AI \cite{AAAI1816669} has shown that when training environments are different from real world data, the trained algorithms fail when applied to real world challenges. To counteract this, we investigate online learning with the DRL algorithm, to allow it to continue learning and adapting as we introduce it to real-world conditions. Our DRL will be trained in a noisy training environment until the DRL agent reaches maximum conversion efficiency. 

The policy is learned which best maps the input space (conversion efficiency and produced noise) to optimum tuning damping factors (actuators), such as to achieve maximum conversion efficiency. DRL is learning-by-doing approach, with trials and multiple episodes, the DRL fine tunes its neural network weights to learn the optimal policy. The final trained DRL agent could provide an optimal transmission signal with efficiency greater than the theoretically optimum.

{\it This is novel research for deep reinforcement learning in quantum space, investigating how online learning can be enabled with DRL algorithms and the influence of neural network architectures on storing quantum noise information and learning solutions quickly.} 

\section{Device Simulation Environment}
\label{sec:sim}

\begin{figure}[t]
	\includegraphics[width = 0.5\textwidth]{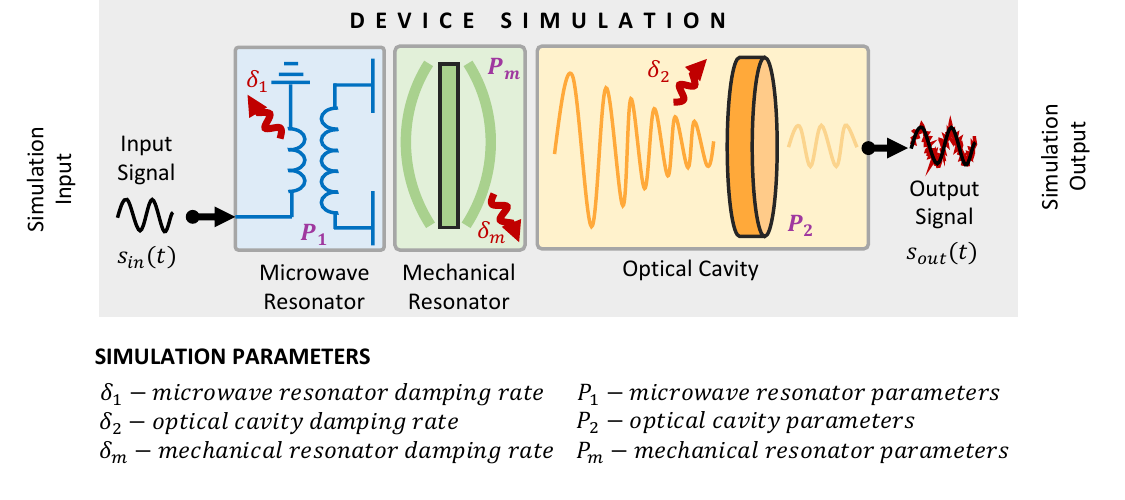}
	\caption{Quantum Transducer Simulation: microwave resonator, mechanical resonator and optical cavity with simulated noise channels.}
	\label{fig:device}
\end{figure}

In order to explore the proposed approach necessary to conduct realistic experiments
and therefore guide our methodology as shown in Figure \ref{fig:methodology} (b), we develop a transducer simulation model. Figure \ref{fig:device} shows an example of the microwave-to-optics transduction device simulation. The example device is composed of three major components: microwave resonator, optical cavity and mechanical resonator that connects them into a single system. The mechanical vibration of the resonator membrane modulates the resonance frequencies of the microwave resonator and optical cavity. Each of the components is described with a set of parameters, i.e. $P_1$, $P_2$ and $P_m$ for microwave resonator, optical cavity and mechanical resonator respectively. With a cycle-accurate simulation, we emulate the device behaviour when converting an input signal $s_{in}(t)$ into an output noisy signal $s_{out}(t)$ using the master equation for open quantum optical systems
\begin{equation}
    \frac{d\rho}{dt} = \mathcal{L}\rho
\end{equation} 

Signal damping rates that are related to the device components, i.e. $\delta_1$, $\delta_2$ and $\delta_m$ for microwave resonator, optical cavity and mechanical resonator respectively are shown as red wavy arrows. By adjusting the simulation parameters within the parameter regime from the NIST experiment and using noise models~\cite{Clerk2010}, we generate a variety of noisy output signals that can be used to train the AI engine towards higher device efficiency. 

\section{Deep Reinforcement Learning}
\label{sec:drl}

\subsection{Algorithm}
In \textit{model-free DRL}, the agents take random decisions within the environment and learn by observing the reward function, overtime they can learn to perform with super-human level ideas. Similar to Monte Carlo Decision Tree search, the neural networks can be designed to learn values or the policies (probabilities) of what action to take in a given state. the optimal policy $\pi ^{*}$ learned is given by

\begin{equation}
    \pi ^{*}= arg\ \underset{\pi}{max}\mathbb{E}[\underset{t\geq 0}{\Sigma}\gamma^{t}r_{t} \mid \pi]
\end{equation}
We design the DRL to perform as a stochastic process (over time) thus the optimal policy allows it to learn actions that will maximize a cumulative reward. 

To build a optimal DRL algorithm for the challenge, we adopt the Actor-Critic methods that allows to model continuous state and action pairs compared to other DRL algorithms. We describe this in Algorithm \ref{alg:drl}. It consists of two parts: the `Critic' that estimates the value function Q, and the `Actor' that updates the policy distribution in the direction suggested by the Critic. 

Developed to optimize multiple parallel agents training and finding optimized solution quickly \cite{pmlr-v48-mniha16}, the algorithm handles environments with noise, diverse domains and aim to achieve equilibrium quicker by introducing more actors to the scenario. In addition to the policy search $\bigtriangledown_{\theta}  J(\theta)$, we use Equation \ref{eg:jtheta} to allow the policy gradient to converge faster thereby minimizing the mean squared error and Bellman update equation. 

\begin{equation}\label{eg:jtheta}
    \bigtriangledown_{\theta}  J(\theta) \sim \Sigma_{t=0}^{T-1}\bigtriangledown_{\theta}log \pi_{\theta} (a_{t}\mid s_{t}) A(s_{t},a_{t})
\end{equation}

\begin{algorithm}
\begin{algorithmic}
\State Initialize parameters $s, \theta, w$ and learning rates $\alpha_{\theta},\alpha_{w}$; sample $a \sim \pi_{\theta}(a|s)$
\For  $\ t=1...T:$ 
     \State Sample reward $r_{t} \sim R(s,a)$ and next state $s' \sim P(s'|s,a)$
     \State Then sample the next action $a' \sim \pi_{\theta}(a'|s')$
     \State Update policy parameters: $\theta \leftarrow \theta + \alpha_{\theta}Q_{w}(s,a)\bigtriangledown_{\theta}log\pi_{\theta}(a|s);$ 
     \State Compute the correction (TD error) for action-value at time $t$:
     \State $\delta_{t}=r_{t}+ \gamma  Q_{w}(s',s')-Q_{w}(s,a)$
     \State and use it to update parameters of Q function:
     \State $w\leftarrow w+\alpha_{w}\delta_{t}\bigtriangledown_{w}Q_{w}(s,a)$
     \State Move to $a\leftarrow a'$ and $s\leftarrow s'$
     \EndFor     
     \end{algorithmic}
\caption{Actor-critic Model}
\label{alg:drl}
\end{algorithm}

\subsection{Online and Adaptive Learning}
The algorithm is trained in a simulated environment described in Section \ref{sec:sim}. However, once trained, the algorithm interacts with real devices to fine tune the neural network weights and learn noise in real experimental setups. For this part, we allow the neural network to perform a 10-episode update every 10 minutes if the accuracy of the neural network drops with time as shown in Figure \ref{fig:methodology} (c).

\section{Conclusions and Future Work}
\label{sec:conc}

Recent reports from the White House and ASCR~\cite{WhitehouseReport, ASCRReport}, both recognize optical-microwave transduction as one of the most challenging and limiting technologies for scaling quantum information technology. Most networking research is focused on extending quantum information transmission distance using quantum repeaters and not optomechanical conversion. Our methodology and early experiments reveal that AI-controlled transduction is imperative for quantum networks. It will enable high-fidelity multi-nodal quantum computation for future quantum information and quantum networking facilities. The resulting software can be integrated with quantum hardware to achieve high efficiencies, not just in simulation, but in real-world experiment and have multi-domain impact across fields of quantum physics.
The software that allows the AI to test and adapt to real quantum device transduction as proof-of-concept to be presented to experimentalists for further research. Our AI tools are deployable and also allow us to investigate explainable models on why the AI will learn certain patterns in these new quantum spaces, producing more data and experiments for physicists and computer scientists to work together and advance the field.

\section*{Acknowledgement}

This work was supported by the Office of Advanced Scientific Computing Research for the Computational Research Division, of the U.S. Department of Energy under Contract No. DE-AC02-05CH11231.

\bibliography{biblio.bib} \bibliographystyle{unsrt}

\end{document}